\documentclass[12pt]{article}
\usepackage{amssymb, amsmath, amsthm}
\usepackage{graphicx}
\usepackage{booktabs}
\usepackage{multirow}

\newcommand{\argmin}{\operatornamewithlimits{arg\,min}}
\usepackage{fullpage}

\newcommand{\mbx}{\boldsymbol{x}}
\newcommand{\mby}{\boldsymbol{y}}

\newcommand{\mbX}{\boldsymbol{X}}
\newcommand{\mbB}{\boldsymbol{B}}

\newcommand{\mbI}{\boldsymbol{I}}

\newcommand{\mbA}{\boldsymbol{A}}

\newcommand{\mbDelta}{\boldsymbol{\Delta}}

\renewcommand{\tilde}{\widetilde}
\renewcommand{\hat}{\widehat}
\newcommand{\mbbeta}{\boldsymbol{\beta}}
\newcommand{\mbsigma}{\boldsymbol{\sigma}}
\newcommand{\mbrho}{\boldsymbol{\rho}}

\newcommand{\mbtheta}{\boldsymbol{\theta}}
\newcommand{\mbTheta}{\boldsymbol{\Theta}}
\newcommand{\mbeta}{\boldsymbol{\eta}}
\usepackage{natbib}
\usepackage{algorithm,algorithmic}
\usepackage{setspace}
\usepackage{blindtext}

\usepackage[table]{xcolor}
\newcolumntype{L}{>$l<$}
\theoremstyle{plain}

\newtheorem{theorem}{Theorem}

\newtheorem{remark}{Remark}

\newtheorem{lemma}{Lemma}

\begin{document}

\title{Heterogeneity-aware integrative regression for ancestry-specific association studies}
\author{Aaron J. Molstad$^*$, Yanwei Cai$^\dagger$, Alexander P. Reiner$^{\dagger, \ddagger}$, \\Charles Kooperberg$^{\dagger, \star}$, Wei Sun$^{\dagger, \star, \mathsection}$, and Li Hsu$^{\dagger, \star}$\medskip\\
School of Statistics, University of Minnesota$^*$\\
Division of Public Health Sciences, Fred Hutchinson Cancer Center$^\dagger$\\
Department of Biostatistics$^\star$ and  Department of Epidemiology$^\ddagger$,\\ University of Washington\\
Department of Biostatistics, University of North Carolina, Chapel Hill$^\mathsection$}
\date{}
\maketitle
\begin{abstract}
Ancestry-specific proteome-wide association studies (PWAS) based on genetically predicted protein expression can reveal complex disease etiology specific to certain ancestral groups. These studies require ancestry-specific models for protein expression as a function of SNP genotypes.  In order to improve protein expression prediction in ancestral populations historically underrepresented in genomic studies, we propose a new penalized maximum likelihood estimator for fitting ancestry-specific joint protein quantitative trait loci models. Our estimator borrows information across ancestral groups, while simultaneously allowing for heterogeneous error variances and regression coefficients. We propose an alternative parameterization of our model which makes the objective function convex and the penalty scale invariant. To improve computational efficiency, we propose an approximate version of our method and study its theoretical properties. Our method provides a substantial improvement in protein expression prediction accuracy in individuals of African ancestry, and in a downstream PWAS analysis, leads to the discovery of multiple associations between protein expression and blood lipid traits in the African ancestry population. \smallskip 
\end{abstract}

\onehalfspacing

\section{Introduction}
\textcolor{black}{Over the last two decades, genomic studies have predominantly sampled from populations of European ancestry} \citep{bentley2017diversity}. Due to genetic differences across ancestries, many of the biological discoveries from these studies are not directly transferable to populations of non-European ancestry. Consequently, the advances in understanding of complex disease etiology may not apply to individuals from underrepresented populations, exacerbating potential health disparities \citep{peprah2015genome}. While there is a growing awareness of the underrepresentation of certain ancestral groups in genomic studies, due to small sample sizes, the power for ancestry-specific analyses remains limited in these groups. Meanwhile, there is evidence that genetic variants associated with complex traits are often shared across ancestral groups \citep{wojcik2019genetic, hou2023causal}. It is thus essential to develop statistical methods that can improve our ability to discover ancestry-specific associations in underrepresented groups, while leveraging commonalities across ancestries. \textcolor{black}{In this work, we use ``ancestry'' as shorthand for ``genetically-inferred ancestry''. }

In this work, we focus on proteome-wide association studies (PWAS) specific to individuals of African ancestry. Proteomics are important because many diseases manifest through changes in protein expression, so proteome-wide association studies can identify novel biomarkers and drug targets \citep{kavallaris2005proteomics}. \textcolor{black}{Although previous studies have demonstrated that the proteome is under genetic control \citep[e.g., see][]{sun2018genomic,robins2021genetic},} relative to the transcriptome, the genetic architecture of the proteome---especially in populations of non-European ancestry---is less well understood due to limited studies \citep{zhang2022plasma}. Our objective is to build a prediction model for protein expression as a function of local/cis SNP genotypes, specific to individuals of African ancestry. We will then use these models to impute  protein expression into large-scale genome-wide association studies (GWAS) and assess their association with complex traits \citep{barbeira2018exploring,dong2020general}. We focus our attention on the potential association between genetically predicted protein expression and five lipid blood traits: low-density lipoprotein cholesterol (LDL), high-density lipoprotein cholesterol (HDL), triglycerides (TG), total cholesterol (TC), and non-high-density lipoprotein cholesterol (nonHDL). It is well known that blood lipid levels are associated with cardiovascular disease risk, and are heritable. Many recent studies have focused on GWAS for lipid levels \citep{graham2021power, 
ramdas2022multi}, though PWAS are very limited \citep{zhang2022plasma}. According to \citet{carnethon2017cardiovascular}, as of 2017, cardiovascular disease was the primary cause of life expectancy differences between African American and White American individuals. Thus, identifying proteins associated with elevated lipid levels may suggest new avenues of research on treating or preventing cardiovascular disease in African American individuals. 

Recent studies suggest that protein expression prediction models do not generalize well across different ancestral groups.  For example, \citet{zhang2022plasma} found that models trained on data collected on individuals of European ancestry (EA) did not perform well when predicting protein expression in individuals of African ancestry (AA). \citet{bhattacharya2022best} found a similar result in the context of gene expression prediction modeling, claiming that ``expression models are not portable across ancestry groups". \textcolor{black}{The results of \citet{patel2022genetic} further suggest that gene expression effect sizes may differ across ancestral groups. } Because of this, \citet{zhang2022plasma} built ancestry-specific proteomic prediction models using data on that ancestral population alone.  However, such an approach does not exploit genetic commonalities between EA and AA populations, and thus may lead to less predictive models. 


While motivated by our desire to build genetic predictive models for proteins in the AA population, our new methodology applies to any regression problem wherein the same responses and predictors are measured on individuals from multiple distinct groups, and the goal is to build a population-specific model leveraging information across populations.
To explain, we give a brief description our proposed method. 
Suppose that for $j \in \{1, \dots, J\} =: [J]$ distinct populations, we observe pairs $\{(y_{(j)i}, \mbx_{(j)i}), i \in [n_{(j)}]\}$ where $n_{(j)}$ is the sample size of the $j$th population. We assume that $y_{(j)i}$ is a realization of the random scalar
\begin{equation} \label{eq:main_model}
 \mbbeta_{*(j)}^\top \mbx_{(j)i} + \epsilon_{(j)i}, ~~~ i \in [n_{(j)}], ~~~ j \in [J],
\end{equation}
where $\mbbeta_{*(j)} \in \mathbb{R}^p$ is the unknown regression coefficient vector for the $j$th population, $\mbx_{(j)i} \in \mathbb{R}^p$ is the predictor for the $i$th subject in the $j$th population, and  $\epsilon_{(j)i} \sim {\rm N}(0, \sigma_{*(j)}^2)$ is random error. In the context of joint pQTL modeling, for example, $y_{(j)i} \in \mathbb{R}$ is a protein's normalized expression and $\mbx_{(j)i} \in \mathbb{R}^p$ are the cis-SNP genotypes for the $i$th subject from $j$th ancestral group. We assume that all $\epsilon_{(j)i}$ are independent and within a population, identically distributed. Notably, we assume that each ancestral population has its own, distinct regression coefficient vector and error variance. \textcolor{black}{Throughout this article, we will refer to a joint pQTL---that is, a SNP whose corresponding coefficient from \eqref{eq:main_model} is nonzero---as a pQTL for simplicity.} 

In previous pQTL studies \citep{zhang2022plasma}, each $\mbbeta_{*(j)}$ was estimated separately using regression techniques designed for large $p$, e.g., elastic net. However, this is less efficient than estimating the matrix $\mbB_* = (\mbbeta_{*(1)}, \dots, \mbbeta_{*(J)}) \in \mathbb{R}^{p \times J}$ directly if there are special structures of $\mbB_*$ that can be exploited. For example, it is reasonable to expect that the majority of pQTLs are shared across multiple populations in the study.  This would imply that the nonzero entries of $\mbB_*$ are concentrated in a small number of rows. Using data from all $J$ populations to identify such rows leads to improved efficiency and parsimony, as our analyses of the ancestral groups in the Women's Health Initiative  (WHI, Section \ref{sec:DataAnalysis}) demonstrate. 

Existing methods for borrowing information across populations often assume homogeneous error variances, e.g., see Section \ref{eq:existing_methods} and references therein. However, we will illustrate that allowing for distinct error variances across populations is important for ancestry-specific pQTL models. This is partly due to the fact that for a SNP genotype, allele frequencies can differ greatly across ancestral groups (e.g., see Figure \ref{fig:LDsubmatrices}). Thus, even if a SNP has the same effect size across populations, the proportion of variability in protein expression explained by the SNP will be affected by the ancestry-specific variability of that SNP. This implies that for a particular protein and set of SNPs, the predictive potential of these SNPs may differ by ancestry. In contrast to many existing methods, our method borrows information across populations while still allowing for heterogeneous error variances.


\section{Methodology}\label{sec:method}
\subsection{Penalized maximum likelihood estimator}
In order to estimate the $\mbbeta_{*(j)}$, we propose a penalized maximum likelihood estimator. Based on the model \eqref{eq:main_model}, the (scaled) negative log-likelihood for the observed data, ignoring constants, is 
$$ L_0(\mbB, \mbsigma) = \frac{1}{2n} \sum_{j=1}^J \sum_{i=1}^{n_{(j)}} \left\{\frac{1}{\sigma_{(j)}^2}(y_{(j)i}- \mbbeta_{(j)}^\top \mbx_{(j)i})^2 + \log(\sigma_{(j)}^2)\right\},$$
where $\mbB = (\mbbeta_{(1)}, \dots, \mbbeta_{(J)}) \in \mathbb{R}^{p \times J}$, $\mbsigma = (\sigma_1, \dots, \sigma_J)^\top \in (0, \infty) \times \cdots \times (0, \infty) =: \mathcal{T}_J,$ and $n = \sum_{j=1}^J n_{(j)}$.
Recall, our key assumption is that the nonzero entries of the $\mbbeta_{*(j)}$ often occur in identical positions, i.e., that pQTLs tend to shared across populations. This would mean that many rows of the matrix $\mbB_*$ are entirely equal to zero, and that among the rows which have nonzero elements, possibly only one element is nonzero. To achieve variable selection of this sort, it is natural to apply the sparse group lasso penalty \citep{simon2013sparse,molstad2021covariance} 
$$ g_{\lambda, \gamma} (\mbB) = \lambda \sum_{l=1}^p \left( \sum_{m=1}^J \mbB_{l,m}^2\right)^{1/2} + \gamma \sum_{l=1}^p \sum_{m=1}^J |\mbB_{l,m}|,$$
where $\lambda>0$ and $\gamma>0$ are a pair of user-specified tuning parameters.
One may thus consider estimating the pair $(\mbB_*, \mbsigma_*)$ using 
\begin{equation}\label{eq:penMLE}
\argmin_{\mbB \in \mathbb{R}^{p \times J}, \mbsigma \in \mathcal{T}_J} \left\{ L_0(\mbB, \mbsigma)  + g_{\lambda, \gamma} (\mbB) \right\}.
\end{equation}
The first term in the penalty function $g$, controlled by tuning parameter $\lambda >0$, encourages row-wise sparsity in estimates of $\mbB_*$.  For $\lambda$ sufficiently large, the solution to \eqref{eq:penMLE} must have some rows entirely equal to zero. The second term in $g$, controlled by tuning parameter $\gamma > 0$, encourages elementwise sparsity in estimates of $\mbB_*$. For example, with a fixed $\lambda$ for $\gamma$ sufficiently large, some rows of the solution to \eqref{eq:penMLE} will have both zero and nonzero entries. 

Despite its appeal, \eqref{eq:penMLE} can perform poorly in practice. The issues are twofold: first, the optimization problem in \eqref{eq:penMLE} is nonconvex. As a consequence, the solution one obtains may be dependent on an algorithm's starting values. The second issue is that the effect of the $\mbbeta_{(j)}$ on the penalty will depend on the $\sigma_{(j)}$, or more generally, the scaling of the $y_{(j)i}.$  If one population has an especially large error variance, then the penalty $g$ will be dominated by its corresponding regression coefficient vectors since reducing the sum of squares in this population will decrease the objective function more overall. Similarly, if the responses have different scalings across populations (e.g., due to measuring the $y_{(j)i}$'s on different platforms, or residualizing using a different set of covariates), then the degree to which each population's regression coefficients are penalized will depend on the response scaling. Populations with larger response scaling will have coefficients penalized more substantially. For these reasons, we recommend against using \eqref{eq:penMLE} in practice. However, by employing a simple reparameterization, both of these issues can be resolved.  
\subsection{Alternative parameterization}
To explain how we avoid these issues, define $\rho_{(j)} = 1/\sigma_{(j)}$ and $\mbtheta_{(j)} = \mbbeta_{(j)}/\sigma_{(j)} = \rho_{(j)} \mbbeta_{(j)}$ for $j \in [J]$. The parameter $\rho_{*(j)}^2 > 0$ is the inverse error variance (i.e., the error precision) and $\mbtheta_{*(j)} \in \mathbb{R}^p$ can be interpreted as the regression coefficient vector for the standardized version of the response (the response with error variance equal to one) in the $j$th population. With this alternative parameterization, we can express the (scaled by $1/n$) negative log-likelihood 
$$ L(\mbTheta, \mbrho) = \frac{1}{2n} \sum_{j=1}^J \sum_{i=1}^{n_{(j)}}\left\{ (y_{(j)i} \rho_{(j)}- \mbtheta_{(j)}^\top \mbx_{(j)i})^2 - \log(\rho_{(j)}^2)\right\},$$
where $\mbTheta = (\mbtheta_{(1)}, \dots, \mbtheta_{(J)}) \in \mathbb{R}^{p \times J}$ and $\mbrho = (\rho_{(1)}, \dots, \rho_{(J)})^\top \in \mathcal{T}_J.$ Since the $\mbtheta_{(j)}$ are on the same scale in the sense that they correspond to standardized predictors and responses, one can impose the penalty $g$ on $\mbTheta$ without the response scalings influencing variables selected. Of course, the sparsity of $\mbtheta_{(j)}$ and $\mbbeta_{(j)}$ will be identical, so penalizing $\mbTheta$ will accomplish the same goal as penalizing $\mbB$ in terms of variable selection.

In view of the preceding discussion, we propose to estimate $\mbB_*$ indirectly using 
\begin{equation}\label{eq:estimator}
\argmin_{\mbTheta \in \mathbb{R}^{p \times J}, \mbrho \in \mathcal{T}} \{ L(\mbTheta, \mbrho) + g_{\lambda, \gamma}(\mbTheta)\}.
\end{equation}
Fortunately, it is easy to show \eqref{eq:estimator} is jointly convex in $\mbTheta$ and $\mbrho$, and biconvex (i.e., convex in $\mbTheta$ with $\mbrho$ held fixed and vice versa). With an estimate of $(\mbTheta_*, \mbrho_*)$ from \eqref{eq:estimator}, say $(\hat\mbTheta, \hat\mbrho)$, we obtain an estimate of $(\mbB_*, \mbsigma_*)$ as $\hat\mbbeta_{(j)} = \hat\mbtheta_{(j)}/\hat\mbrho_{(j)}$ and $\hat\sigma_{(j)} = 1/\hat\rho_{(j)}$ for $j \in [J].$

This reparameterization of the normal linear regression model was originally proposed by \citet{stadler2010L1} in the context of mixture modelling. Specifically, \citet{stadler2010L1} proposed a version of \eqref{eq:estimator} with $\lambda = 0$ and the populations from which each observation was drawn unknown. \citet{khalili2013regularization} generalized the approach of \citet{stadler2010L1} to handle more general penalty functions (e.g., SCAD and MCP). Recently, \citet{li2021pursuing} also used this reparameterization in the context of mixture modelling where they decomposed each $\mbbeta_{(j)}$ into the sum of $\mbeta_0$ and $\mbeta_{(j)}$, i.e., $\mbbeta_{(j)} = \mbeta_0 + \mbeta_{(j)}$ where $\mbeta_0$ are effects common to all populations and $\mbeta_{(j)}$ are effects specific to the $j$th population. 

For the sake of space, we describe how to solve the optimization problem in \eqref{eq:estimator} in the Supplementary Material. To summarize, we use a blockwise coordinate descent algorithm (Algorithm 1) that alternates between updating $\mbrho$ with $\mbTheta$ fixed, and vice versa. 

\subsection{Relation to existing approaches}\label{eq:existing_methods}
The estimator \eqref{eq:estimator} is related to a number of existing methods for integrative regression \citep[see ``horizontal integration" in reivew article of ][]{richardson2016statistical} One type of integrative regression analysis is where data collected on distinct populations are analyzed jointly in order to exploit or discover similarities or differences between populations. A subset of such methods are described in the review of \citet{zhao2015integrative}. The methods described therein are characterized in terms of exploiting either the ``homogeneity model'', where it is assumed that either the $k$th row of $\mbB_*$ is entirely zero, $\mbB_{*k,\cdot} = 0$ or entirely nonzero; or the ``heterogeneity model'', where some elements of a particular row may be zero and others nonzero. As described in \citet{zhao2015integrative}, the heterogeneity model is especially useful when modeling distinct populations. However, none of the method described in \citet{zhao2015integrative} are able to account for distinct variances/response scaling in each population. For example, the method of \citet{zhao2015integrative}, with MCP penalty replaced with the scaled identity function, corresponds to \eqref{eq:estimator} with $\gamma = 0$ under the additional restriction that $\sigma_{*(j)} = \sigma_{*(k)}$ for all $j \neq k$. As we will show in our simulation studies, when $\sigma_{*(j)}$ and $\sigma_{*(k)}$ are distinct, this method can perform poorly. Even when $\sigma_{*(j)} = \sigma_{*(k)}$, in general, our method still performs as well as that which correctly assumes they are equal.

Another class of methods assumes that $\mbbeta_{*(j)} \approx \mbbeta_{*(k)}$ in the sense that the distance between $\mbbeta_{*(j)}$ and $\mbbeta_{*(k)}$ is small (e.g., in squared Euclidean or $L_1$-norm distance). For example, one could use the variation of the fused lasso penalty \citep{tibshirani2005sparsity,tang2016fused} such as $\sum_{l=1}^p \sum_{j \neq k} |\mbbeta_{(j)l} - \mbbeta_{(k)l}|$, which would encourage estimates of $\mbbeta_{*(j)}$ and $\mbbeta_{*(k)}$ with some entries exactly equivalent to one another. Alternatively, one could use a penalty which fuses coefficients which are thought to be similar to one another based on some initial estimate \citep{wang2016fused}. These approaches are  useful in applications where one can expect regression coefficients to be same across populations---for example, when the distinct datasets represent data collected on the same population in different studies---but in our context, may inappropriately bias our estimates towards a model assuming that $\mbbeta_{*(j)} = \mbbeta_{*(k)}$. Discovering differences in effect sizes is one of the primary objectives of our study, so this type of bias is especially undesirable. Moreover, this type of penalty is not invariant to response scaling, so it would suffer from the same issues as \eqref{eq:penMLE}. 

\textcolor{black}{
Our work is also related to recent literature on transfer learning \citep{weiss2016survey,suder2023bayesian}. Transfer learning for linear regression often assumes the same data generating model as \eqref{eq:main_model}, but is focused on coefficients from one target population, say, $\mbbeta_{*(0)}.$ Often, methods for transfer learning assume $\|\mbbeta_{*(0)} - \mbbeta_{*(j)}\|_2$ is small \citep{li2022transfer}, or that the angle between $\mbbeta_{*(0)}$ and $\mbbeta_{*(j)}$ is small \citep{gu2022robust}. However, in the work on transfer learning, it is often assumed that all datasets cannot be accessed by a analyst user simultaneously, whereas we assume all $J$ datasets are on hand for model building. }

\textcolor{black}{
Finally, our method is also related to existing methods for estimation of polygenic risk scores. For example, in the context of cross-population polygenic prediction, \citet{ruan2022improving} assumes exactly the model \eqref{eq:main_model} (where their response is an arbitrary phenotype) but in a Bayesian framework. Like us, they devise a shrinkage estimator of the regression coefficients that shares information across populations.  However, much of the work on polygenic prediction assumes the individual level data are not available---only summary statistics due to the large sample size of genome-wide association studies \citep{knight2024multi}. Devising a variation of our method applicable with only summary statistics is beyond the scope of the present article, but a reasonable starting point would build on the approach from \citet{mak2017polygenic}.}

\section{Accommodating ancestry-specific SNPs}\label{sec:AncestrySpecific}
As mentioned, genetic variation differs across ancestral groups. For example, populations of African ancestry are characterized by greater genetic diversity and less linkage disequilibrium than  those of European ancestry \citep{campbell2008african}.
Thus, it is often the case that the columns of 
$\mbX_{(j)}$ and $\mbX_{(k)}$, for example, consist of overlapping, though distinct, sets of SNPs. Our method can be easily modified to accommodate this setting, which we detail in this section. In this context, the goal is still to model the conditional distribution of protein expression given SNP genotypes, however, the SNP genotypes conditioned on may be distinct across the $J$ populations in the study. It is arguably more natural to assume that error variances will differ across populations in this case.

First, let $\mathcal{A}(j)$ denote the set of SNPs available for the $j$th population and let $\mathcal{A} = \cup_{j=1}^J \mathcal{A}(j)$ be the set of all SNPs available in at least one study. In order to exploit shared pQTLs, we propose a slight modification of our penalty function. For the moment, define $\mbTheta \in \mathbb{R}^{p^* \times J}$ where $p^*$ is the cardinality of $\mathcal{A}$. Let $\mathcal{B}(k)$ denote the index set for the populations from which the $k$th SNP (i.e., $k$th element of $\mathcal{A}$) is available for $k \in [p_*].$ Then, define
$$g_{\lambda, \gamma}^{\mathcal{B}}(\mbTheta) = \lambda \sum_{k=1}^p \sqrt{|\mathcal{B}(k)|} \|\mbTheta_{k,\mathcal{B}(k)}\|_2 + \gamma \sum_{k=1}^p \sum_{j \in \mathcal{B}(k)} |\mbTheta_{j,k}|$$
where $\mbTheta_{k,\mathcal{B}(k)} \in \mathbb{R}^{|\mathcal{B}(k)|}$ is the subvector of $\mbTheta_{k,\cdot} \in \mathbb{R}^{J}$ consisting only of the components whose indices belong to $\mathcal{B}(k)$. To estimate the regression coefficients, we then define the estimator
\begin{equation}\label{eq:mismatchEst}
\argmin_{\mbTheta \in \mathbb{R}^{p \times J}, \mbrho \in \mathcal{T}_J} \left[ \frac{1}{2n} \sum_{j=1}^J\left\{ \|\mby_{(j)} \rho_{(j)}-  \mbX_{(j)}\mbTheta_{\mathcal{A}(j),j}\|_2^2 - n_{(j)}\log(\rho_{(j)}^2)\right\} + g^{\mathcal{B}}_{\lambda, \gamma}(\mbTheta)\right], 
\end{equation}
$$\text{subject to}~~ \mbTheta_{\mathcal{A}\setminus \mathcal{A}(j),j} = 0~~ \forall j \in [J].$$
The additional constraint that $\mbTheta_{\mathcal{A}\setminus \mathcal{A}(j),j} = 0$ requires that coefficients for the SNPs not included in the $j$th population's SNP set must be equal to zero. This reflects that these SNPs' effects are not estimable with the data at hand.  Just as before, the objective function in \eqref{eq:mismatchEst} is convex in the pair ($\mbTheta, \mbrho$) and the algorithm used to solve \eqref{eq:estimator} can be applied to compute \eqref{eq:mismatchEst} with minor modification. 
\begin{remark}
Let $p^*$ be the number of distinct SNPs across all $J$ populations in a study. 
The solution to \eqref{eq:mismatchEst} can be obtained by solving a (weighted) version of \eqref{eq:estimator} where the genotype matrices $\mbX_{(j)}$ are replaced with $\tilde{\mbX}_{(j)} \in \mathbb{R}^{n_{(j)} \times p^*}$, where $\tilde{\mbX}_{(j)}$ is a padded version of $\mbX_{(j)}$ with columns of zeros added for all SNPs not recorded in the $j$th population.
\end{remark}
The preceding remark establishes that computing \eqref{eq:mismatchEst} does not require a new computational procedure. This remark follows from the fact that if a column of $\tilde{\mbX}_{(j)}$ is entirely zero, then making the corresponding entry of $\mbtheta_{(j)}$ nonzero can only increase the objective function. 
\section{Fast two-step approximation}\label{sec:Approx}
\subsection{Algorithm}
Though \eqref{eq:estimator} is well-motivated, in practice it can require more computation time than existing estimators due to the iterative nature of Algorithm 1. In Section \ref{sec:Theory} and \ref{sec:SimulationStudies}, we show both theoretically and empirically that a simple two-step approximation to \eqref{eq:estimator} can work as well as \eqref{eq:estimator} at substantially lower computational cost.  In particular, this approach requires first obtaining a pilot estimator of the $\sigma_{*(j)}$, fixing this as our estimate of $\rho_*$, and then solving the optimization problem in \eqref{eq:estimator} with respect to $\mbTheta$ alone. 
Two-step approximations to the joint normal maximum likelihood estimators of both mean and variances is a common in the high-dimensional literature, e.g., see \citet{rothman2010sparse} or \citet{Li2020highdimensional}

Formally, the two-step approximation algorithm proceeds as follows. 
\begin{enumerate}
    \item[1.] For each $j \in [J]$ separately, compute 
    \begin{equation}\label{eq:naturalLasso} \check{\sigma}_{(j)}^2 = \min_{\mbbeta_{(j)} \in \mathbb{R}^p}\left(\frac{1}{2n_{(j)}}\|\mby_{(j)} - \mbX_{(j)}\mbbeta_{(j)}\|_2^2 + \phi_{(j)} \|\mbbeta_{(j)}\|_1\right)
    \end{equation}
    with tuning parameter $\phi_{(j)} > 0$ chosen to minimize the cross-validated prediction error. 
    \item[2.1.] With $\check{\mbrho} = (1/\check{\sigma}_{(1)}, \dots, 1/\check{\sigma}_{(J)})^\top$ fixed, compute 
    \begin{equation}\label{eq:ApproxEstimator}
    \check{\mbTheta} \in \argmin_{\mbTheta \in \mathbb{R}^{p \times J}} \{ L(\mbTheta, \check{\mbrho}) + g_{\lambda, \gamma}(\mbTheta)\},
    \end{equation}
    where $\lambda> 0$ and $\gamma > 0$ are chosen by cross-validation. 
    \item[2.2.] For each $j \in [J]$, return estimate of $ \check{\mbtheta}_{(j)}\check{\sigma}_{(j)}$ as an estimate of $\mbbeta_{*(j)}$.
\end{enumerate}

Pilot estimation of the $\sigma_{*(j)}$, Step 1, could be done using any method designed for variance estimation in high-dimensional linear models, e.g., see the review of \citet{reid2016study}. We use the so-called natural lasso estimator \citep{yu2019estimating}, defined as in \eqref{eq:naturalLasso}. 
This estimator of $\sigma_{*(j)}$ is computationally efficient as it requires only solving an $L_1$-penalized least squares problem, and as shown in \citet{yu2019estimating}, performs well across diverse settings.

Given $\check{\mbrho} = (1/\check{\sigma}_{(1)}, \dots, 1/\check{\sigma}_{(J)})^\top$, in Step 2 we estimate $\mbbeta_{*(j)}$ indirectly using $\check{\mbtheta}_{(j)}\check{\sigma}_{(j)}$. Solving \eqref{eq:ApproxEstimator} is significantly easier than solving \eqref{eq:estimator} since we need not ever update $\mbrho$. In this case, \eqref{eq:ApproxEstimator} can solved by applying an accelerated variation of Algorithm 1 \citep[Chapter 4.3]{parikh2014proximal} wherein Step 1 is skipped.

\subsection{Statistical properties}\label{sec:Theory}
We study the theoretical properties of the approximate version of our estimator with both $n, p,$ and $J$ growing. We implicitly treat both $p$ and $J$, which can both diverge, as functions of $n.$ We consider the case that the error variances $\sigma_{*(j)}^2$ are estimated by generic pilot estimator $\tilde{\sigma}_{(j)}^2$. While these results are not directly applicable to the exact version of our method, the theoretical properties of this estimator provide useful insights regarding how selecting important SNPs across populations can improve estimation accuracy. The estimator of $\mbTheta_*$ we study, $\widetilde{\mbTheta}$, is thus a version of \eqref{eq:estimator} with $\gamma = 0$, 
$$\argmin_{\mbTheta \in \mathbb{R}^{p \times J}} \{L(\mbTheta, \tilde{\mbrho}) + g_{\lambda, 0}(\mbTheta)\},$$
where $\tilde{\mbrho}$ is the vector $(1/\tilde{\sigma}_{(1)}, \dots, 1/\tilde{\sigma}_{(J)})^\top$ of pilot inverse standard deviations. Similar results can be established with $\gamma > 0$, but this would complicate our notation and assumptions. 

We first need to define a number of important quantities and assumptions. Let $\mathcal{S} = \{k: \mbB_{*k,\cdot} \neq 0, k \in [p]\}$ be the set of indices of SNPs which are important in at least one population and let $\mathcal{S}^c = [p]\setminus \mathcal{S}$. Let the notation $\mbA_{\mathcal{S}}$ be the projection of the matrix $\mbA$ onto the set of matrices with sparsity pattern defined by $\mathcal{S}$, i.e., $\mbA_{\mathcal{S}} = \argmin_{\mbB \in \mathbb{M}(\mathcal{S})} \|\mbA - \mbB\|_F^2$ where $\mathbb{M}(\mathcal{S}) = \{\mbB \in \mathbb{R}^{p \times J}: \mbB_{k,\cdot} = 0 ~~ \text{for all } k \in \mathcal{S}^c\}$ and $\|\cdot\|_F$ is the Frobenius norm. Next, for fixed constant $M \geq 0$, define the set 
$\mathbb{C}(\mathcal{S}, M) = \left\{\mbDelta \in \mathbb{R}^{p \times J}: \|\mbDelta_{\mathcal{S}^c}\|_{1,2} \leq 3\|\mbDelta_{\mathcal{S}}\|_{1,2} + M \|\mbDelta\|_F \right\},$
where $\|\mbA\|_{1,2} := \sum_{j} \|\mbA_{j,\cdot}\|_2$. For relatively small $M$, $\mathbb{C}(\mathcal{S}, M)$ consists of matrices whose row-wise Euclidean norms are larger in rows indexed by $\mathcal{S}$ relative to $\mathcal{S}^c$. As $M$ increases, row-wise norms in $\mathcal{S}^c$ can become larger. Loosely speaking, larger $M$ means $\mathbb{C}(\mathcal{S}, M)$ has more volume and vice versa.


With these quantities in hand, we can now characterize the conditions and assumptions we need in order to establish our error bound. 
\begin{itemize}
  \item \textbf{C1.} (Blockwise standardization) For each $k \in [p]$, the columns of each $\mbX_{(1)}, \dots, \mbX_{(J)}$ are scaled such that $\|[\mbX_{(j)}]_{\cdot, k}\|_2^2 \leq n_{(j)}$.
\end{itemize}
The condition $\textbf{C1}$ is a common standardization condition required for lasso-type estimators. This condition is a matter of scaling the columns of each $\mbX_{(j)}$, which we treat as fixed. Next, we state our needed assumptions. 
\begin{itemize}
  \item \textbf{A1.} (Normal errors) Each $\mby_{(j)}$ is a realization of $\mbX_{(j)}\mbbeta_{*(j)} + \boldsymbol{\epsilon}_{(j)}$ where the random vector $\boldsymbol{\epsilon}_{(j)} \sim {\rm N}_{n_{(j)}}(0, \sigma_{*(j)}^2 \mbI_{n_{(j)}})$ independently for all $j \in [J]$. 
    \item \textbf{A2.} (Restricted eigenvalue condition) For a given $\mathcal{S}$ and positive constant $M$, there exists a constant $k$ such that $0 < k \leq \kappa_M(\mathcal{S})$ where 
  $$ \kappa_M(\mathcal{S}) = \inf_{\mbDelta \in \mathbb{C}(\mathcal{S},M), \|\mbDelta\|_F^2 = 1} \sum_{j=1}^J \frac{\|\mbX_{(j)} \mbDelta_{(j)}\|_2^2}{2n_{(j)}}.$$
   \item \textbf{A3.} (Nonvanishing sample size) There exists positive constants $\underline{\pi}$ and $\bar{\pi}$ such that 
  $0 < \underline{\pi} \leq  \min_{j \in [J]}
n_{(j)}/n \leq \max_{j \in [J]} n_{(j)}/n \leq \bar\pi < 1$.
\end{itemize}
Assumption \textbf{A1} requires that the errors for each population are normally distributed, i.e., that \eqref{eq:main_model} holds. 
Assumption $\textbf{A2}$ is a generalization of the standard restricted eigenvalue condition \citep{raskutti2010restricted,negahban2012unified}. In particular, if $J = 1$ and $M = 0$, this would be exactly the standard restricted eigenvalue condition, which assumes the existence of a positive constant $\kappa$ such that 
$\|\mbX_{(j)} \mbDelta_{(j)}\|_2^2/(2n_{(j)}) \geq \kappa \|\mbDelta_{(j)}\|_2^2$ for all $\mbDelta_{(j)}$ in a restrictive set like $\mathbb{C}(\mathcal{S}, 0).$ Our restricted eigenvalue, in contrast, involves the summation over each of the $J$ populations. Note that the larger $M$ is, the smaller $\kappa_M(\mathcal{S})$ will be in general. 

Assumption $\textbf{A3}$ requires that as the total sample size $n = \sum_{j=1}^J n_{(j)}$ grows, each of the populations' corresponding sample sizes $n_{(j)}$ grow at a nonvanishing rate.  If the population to sample from is drawn randomly, this means that the probability of drawing from population $j$ is bounded below for $j \in [J]$, whereas if the population to sample from is chosen deterministically, no populations are systemically undersampled. 

Finally, given our pilot estimates $\tilde{\mbrho}$, we define $R_n(\tilde{\mbrho})$, as 
$$ R_n(\tilde{\mbrho}) =  \frac{1}{n^2}\sum_{j=1}^J \left(1 - \frac{\sigma_{*(j)}}{\tilde{\sigma}_{(j)}}\right)^2 \frac{\|\mbX_{(j)}^\top \mbX_{(j)}\mbbeta_{*(j)}\|_2^2}{\sigma_{*(j)}^2}.$$
The quantity $R_n(\tilde{\mbrho})$ is needed to characterize the additional error incurred from using a pilot estimate of $\sigma_{*(j)}$, $\tilde{\sigma}_{(j)}$, in our estimation criterion. Intuitively, if $\tilde{\sigma}_{(j)} \to \sigma_{*(j)}$ sufficiently fast, then $R_n(\tilde{\mbrho}) \to 0$.

We are now ready to state our key technical result.  
\begin{lemma}\label{thm1}
If $\lambda = 2\sqrt{\bar{\pi}} \max_{j \in [J]} (\sigma_{*(j)}/\tilde{\sigma}_{(j)})[(J/n)^{1/2} + \{2k_1 \log (p)/n\}^{1/2}]$ for fixed constant $k_1 > 1$, there exists a constant $M > 0$ such that $\sqrt{R_n(\tilde{\mbrho})} \leq \lambda M /2$, and \textbf{C1}, \textbf{A1}--\textbf{A3} hold, then 
$$ \|\widetilde{\mbTheta} - \mbTheta_*\|_F^2 \leq  \frac{32|\mathcal{S}|}{\kappa^2_{M}(\mathcal{S})} \left(\frac{\bar\pi}{\underline{\pi}^2}\right) \left\{\max_{j \in [J]} \frac{\sigma^2_{*(j)}}{\tilde{\sigma}^2_{(j)}} \left(\frac{J + 2 k_1 \log (p)}{n} \right)  + \frac{R_n(\tilde{\mbrho})}{3 \bar{\pi}|\mathcal{S}|} \right\}$$
with probability at least $1 - p^{1-k_1}$.
\end{lemma}
\textcolor{black}{
The result of Lemma \ref{thm1} demonstrates how various dimensions of the problem affect estimation accuracy of $\mbTheta_*$. In particular, we can see that the dimension of the candidate SNP set, $p$, only affects the estimation error logarithmically. The $|\mathcal{S}|\log (p)/n$ term is standard when using penalized least squares estimators to perform variable selection in \eqref{eq:main_model} \citep{negahban2012unified}. Similarly, $J$, the number of populations in our study, appears in the bound through $|\mathcal{S}|J$, which is the number of coefficients in rows of $\mbB_*$ with at least one nonzero entry.  }

\textcolor{black}{Lemma \ref{thm1} also illustrates the interactions between the statistical error and the approximation error. The approximation error appears both in the term $R_n(\tilde{\tilde{\mbrho}})$, and in the restricted eigenvalue. If $R_n(\tilde{\tilde{\mbrho}})$ is large, then $M$ must be large,  which will diminish $\kappa_{M}(\mathcal{S})$. The first term in the sum on the right hand side (the statistical error) would, at first glance, seem to suggest that overestimating $\sigma_{*(j)}^2$ is beneficial (since then, $\max_{j \in [J]} \sigma^2_{*(j)}/\tilde{\sigma}_{(j)}^2$ would be small).  This, however, has the effect of inflating $R_n(\tilde{\mbrho})$ (relative to $\tilde{\sigma}_{(j)}$ such that $\sigma_{*(j)}/\tilde{\sigma}_{(j)}$ is close to one). Moreover, as our main result will illustrate, this would have a negative effect on the estimation accuracy of $\mbB_*.$  
\begin{theorem}\label{lemma1}
Define the two-step estimator of $\mbbeta_{*(j)}$, $\widetilde{\mbbeta}_{(j)} = \widetilde{\mbtheta}_{(j)} \tilde{\sigma}_{(j)}$ for $j \in [J]$ and let $\widetilde{\mbB} = (\widetilde\mbbeta_{(1)}, \dots, \widetilde\mbbeta_{(J)})$. Let $\tilde{\sigma}_{\max} = \max_{j \in [J]} \tilde{\sigma}_{(j)}$ .
If the conditions of Lemma \ref{thm1} hold, then
\begin{align*}
& \|\widetilde\mbB - \mbB_*\|_F^2 \leq \\
& \left[  \frac{64\tilde{\sigma}_{\max}^2 |\mathcal{S}|}{\kappa^2_{M}(\mathcal{S})} \left(\frac{\bar\pi}{\underline{\pi}^2}\right) \left\{\max_{j \in [J]} \frac{\sigma^2_{*(j)}}{\tilde{\sigma}^2_{(j)}} \left(\frac{J + 2k_1 \log (p)}{n} \right)  + \frac{R_n(\tilde{\mbrho})}{3\bar{\pi}|\mathcal{S}|} \right\} + 2\sum_{j=1}^J \left(1 - \frac{\tilde\sigma_{(j)}}{\sigma_{*(j)}}\right)^2\|\mbbeta_{*(j)}\|_2^2\right]
 \end{align*}
 with probability at least $1 - p^{1-k_1}.$
\end{theorem}
Theorem \ref{lemma1} illustrates that when estimating $\mbB_*$ using the two step approximation, there is additional approximation error---beyond that incurred in the estimation of $\mbTheta_*$---given by $\sum_{j=1}^J \left(1 - \tilde\sigma_{(j)}/\sigma_{*(j)}\right)^2\|\mbbeta_{*(j)}\|_2^2.$ However, if our pilot estimators of the $\sigma_{*(j)}$ converge sufficiently fast in the sense that $\sum_{j=1}^J (1 -  \tilde{\sigma}_{(j)}/\sigma_{*(j)})^2 \max\{1, \|\mbbeta_{*(j)}\|_2^2\} = o(\lambda^2)$ and $R_n(\tilde{\mbrho}) = o(\lambda^2)$, then $\|\widetilde\mbB - \mbB_*\|_F^2$ will be dominated by the statistical error from estimating $\|\widetilde\mbTheta - \mbTheta_*\|_F^2$ asymptotically.}


\section{Simulation studies}\label{sec:SimulationStudies}

\subsection{Data generating models and competing methods}
In this section, we illustrate the performance of our method through simulation studies. Because our motivating application---described in the Introduction---focuses on the case that $J = 2$, we restrict our attention to the $J=2$ version of method throughout this section. Our simulation studies are intended to demonstrate how the number of shared pQTLs, the ratio $\sigma_{*({\rm AA})}/\sigma_{*({\rm EA})}$, and the signal-to-noise ratio affect the performance of our method versus a number of relevant competitors. 

\begin{figure}
\begin{center}
\includegraphics[width=0.5\textwidth]{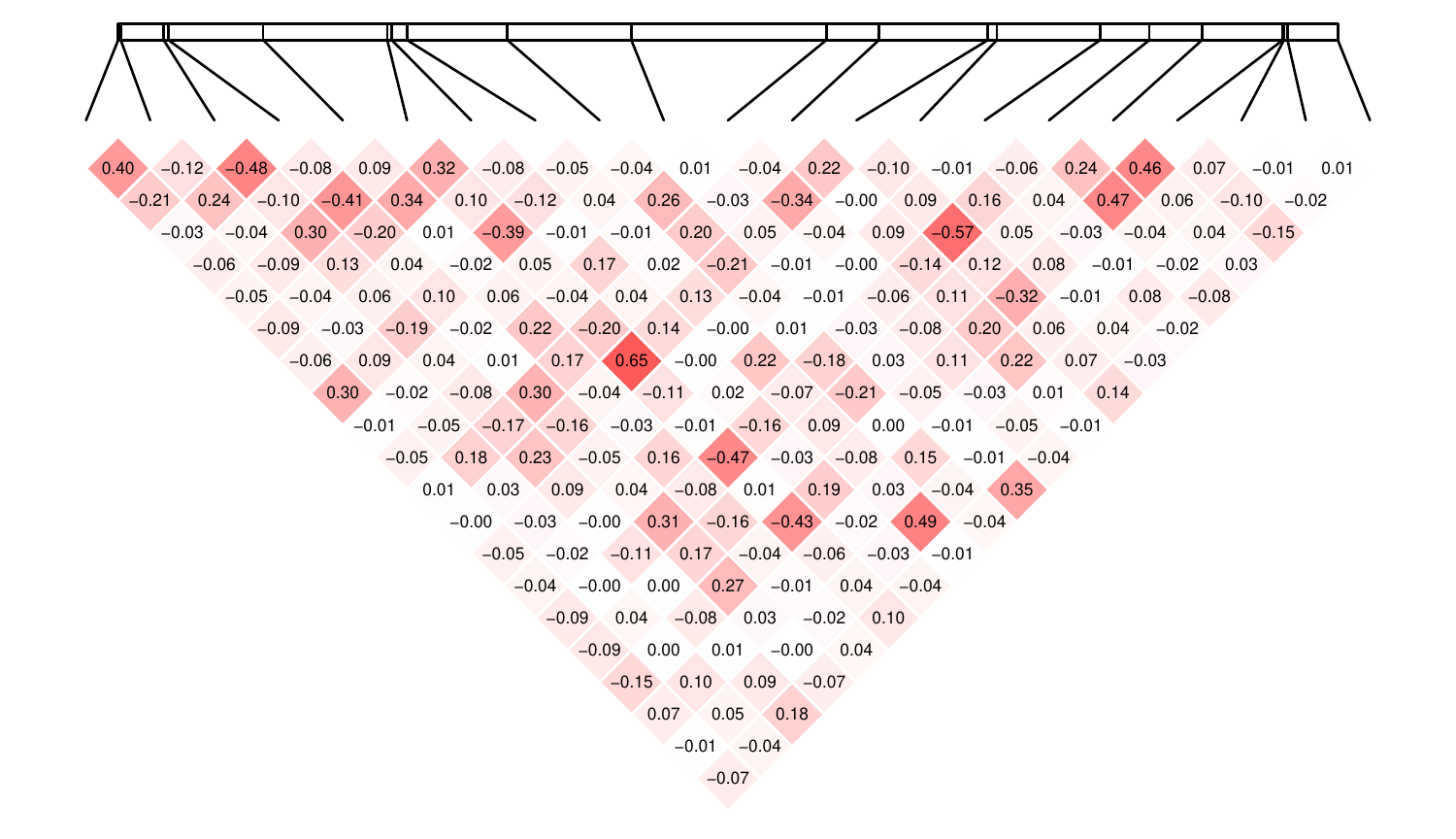}\includegraphics[width=0.5\textwidth]{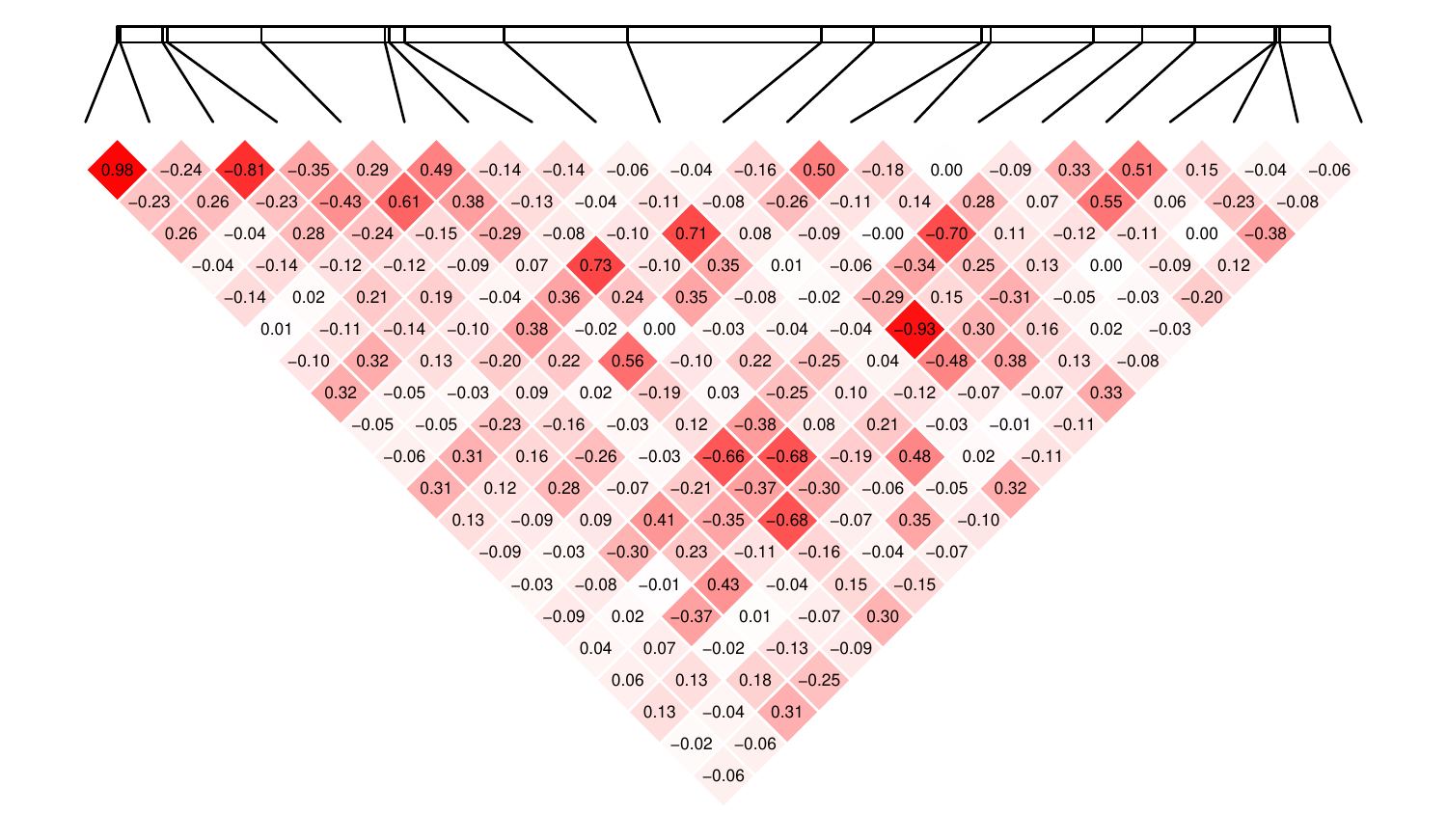}
\hfill(A) African ancestry \hspace{120pt} (B) European ancestry
\end{center}
\caption{Linkage disequilibrium matrices for subjects of African ancestry and subjects of European ancestry corresponding to twenty consecutive SNPs from Chromosome 19, 35103145--35119821. Darker cells denote larger correlations in magnitude.}\label{fig:LDsubmatrices}
\end{figure}

To closely mimic the settings of our motivating data analysis, we use real genotype data collected on individuals of both African ancestry (AA), 1033, and European ancestry (EA), 866. \textcolor{black}{These are a subset of the real genotype data analyzed in Section \ref{sec:DataAnalysis} and their source is described in Section \ref{subsec:dataPrep}.} These SNPs are from Chromosome 19, 34759785---36716111. In Figure \ref{fig:LDsubmatrices} we display linkage disequilibrium (LD) submatrices for both EA and AA populations for the particular set of SNPs we use. Evidently, the LD structure between populations is fundamentally distinct: in the EA population, there are much stronger positive correlations between SNPs in the leftmost part of the figures, whereas in AA, the correlations are somewhat closer to zero. This coheres with well known differences in genetic architecture between these two ancestral groups \citep{campbell2008african}. 

\textcolor{black}{Given the $p = 776$ SNP genotypes retained after preprocessing (see Section \ref{subsec:dataPrep})}, for 100 independent replications under each model setting, we generated protein expression independently for individuals from both populations. 
First, we randomly generated $\mbbeta_{*({\rm EA})}$ and $\mbbeta_{*({\rm AA})}$ to each have $40$ nonzero entries. Given $q \in [0,1],$ we first choose $40 q$ predictors to be nonzero in both coefficients, then independently choose $40  (1-q)$ predictors to be nonzero in each of the two population respectively, where $q$ is the proportion of shared pQTLs. 

We set all nonzero entries of both $\mbbeta_{*({\rm EA})}$ and $\mbbeta_{*({\rm AA})}$ to have magnitude equal to the inverse of the corresponding SNP's standard deviation within the respective population. Each nonzero entry is positive with probability 0.8 or negative with probability 0.2. \textcolor{black}{This choice (0.8/0.2) reduces the probability that shared pQTLs have different signs across populations. } Next, we determine the value of $\sigma_{*({\rm EA})}$ in order to fix the signal-to-noise ratio (SNR, equal to $R^2/(1- R^2)$), in the EA population. Specifically, we set $\sigma_{*({\rm EA})}$ to be $\sqrt{{\rm SNR}}$ times the standard deviation of $\mbX_{({\rm EA})}\mbbeta_{*({\rm EA})}$. Then, we determine $\sigma_{*({\rm AA})}$ according to the ratio $\sigma_{*({\rm AA})}/\sigma_{*({\rm EA})}$, which varies across simulation settings.  Finally, $\mby_{(\rm AA)}$ and $\mby_{({\rm EA})}$ are generated from \eqref{eq:main_model} (with distinct intercepts included).  We consider ${\rm SNR} \in \{1/2, 1, 2\}$, $q \in \{0.6, 0.8, 0.9\}$, and $\sigma_{*({\rm AA})}/\sigma_{*({\rm EA})} \in \{1, 1.25, 1.5, 2, 3\}.$ When $\sigma_{*({\rm AA})}/\sigma_{*({\rm EA})} > 1$, the SNR in the AA population can be substantially lower than that in the EA population.  To assess the affect of the sample size, we perform simulations with $n_{\rm (AA)} = 1033$ and $n_{\rm (AA)} = n_{\rm EA}/2 = 433.$ When using the reduced AA sample size, we randomly sample AA individuals independently in each replication. 

We focus our attention on estimation performance for the parameters of the AA population. 
We consider multiple competing methods. 
The first competitor is the separate elastic net (\texttt{SEN}) estimator, defined as 
\begin{equation} \label{eq:sepEN}
\argmin_{\mbbeta \in \mathbb{R}^p} \left\{ \frac{1}{2n_{({\rm AA})}} \|\mby_{({\rm AA})} - \mbX_{({\rm AA})} \mbbeta\|_2^2 + \alpha \lambda \|\mbbeta\|_1 + \frac{\lambda(1-\alpha)}{2}\|\mbbeta\|^2_2 \right\}
\end{equation}
where the pair of tuning parameters $(\lambda, \alpha) \in (0,\infty) \times [0,1]$ are chosen using ten-fold cross-validation.  The second competitor is what we call the ``agnostic'' elastic net estimator (\texttt{AEN}), which ignores that the data come from two distinct populations. Namely, this estimator stacks $\mby_{({\rm both})} = (\mby_{(\rm AA)}^\top,\mby_{(\rm EA)}^\top)^\top$ and $\mbX_{(\rm both)} = (\mbX_{(\rm AA)}^\top,\mbX_{(\rm EA)}^\top)^\top$, and solves \eqref{eq:sepEN} for input response $\mby_{({\rm both})}$ and predictor matrix $\mbX_{(\rm both)}$ (though allows for distinct intercepts between the two populations). Note that \texttt{AEN} is the estimator which implicitly assumes both $\mbbeta_{*(\rm AA)} = \mbbeta_{*(\rm EA)}$ and $\sigma_{*({\rm AA})} = \sigma_{*({\rm EA})}$

The other competitors we consider are all special cases of our method. The exact version of our method, \eqref{eq:estimator}, we call \texttt{HEAT} for \underline{he}terogeneity-\underline{a}ware join\underline{t} estimator. We also consider the two-step approximation to \eqref{eq:estimator} described in Section \ref{sec:Approx}, denoted \texttt{HEAT-App}, and an oracle version of \texttt{HEAT}, \texttt{HEAT-Orc}, which fixes $\mbrho$ at the truth $\mbrho_*$. By comparing to \texttt{Heat-Orc}, we can see how estimation accuracy is affected by estimating variances in \eqref{eq:estimator}.  Finally, another reasonable estimator is that which assumes $\sigma_{*({\rm EA})} = \sigma_{*({\rm AA})},$ but exploits shared sparsity. Such an estimator (\texttt{reHEAT} for ``restricted HEAT") can be characterized as \eqref{eq:penMLE} with $\sigma_{(\rm AA)} = \sigma_{(\rm EA)}$ both fixed and set equal to one. This corresponds to the penalized maximum likelihood estimator of $\mbB_*$ from \eqref{eq:penMLE} for any value of the variances, assuming they are equivalent. 

\begin{figure}[t]
\begin{center}
\includegraphics[width=0.9\textwidth]{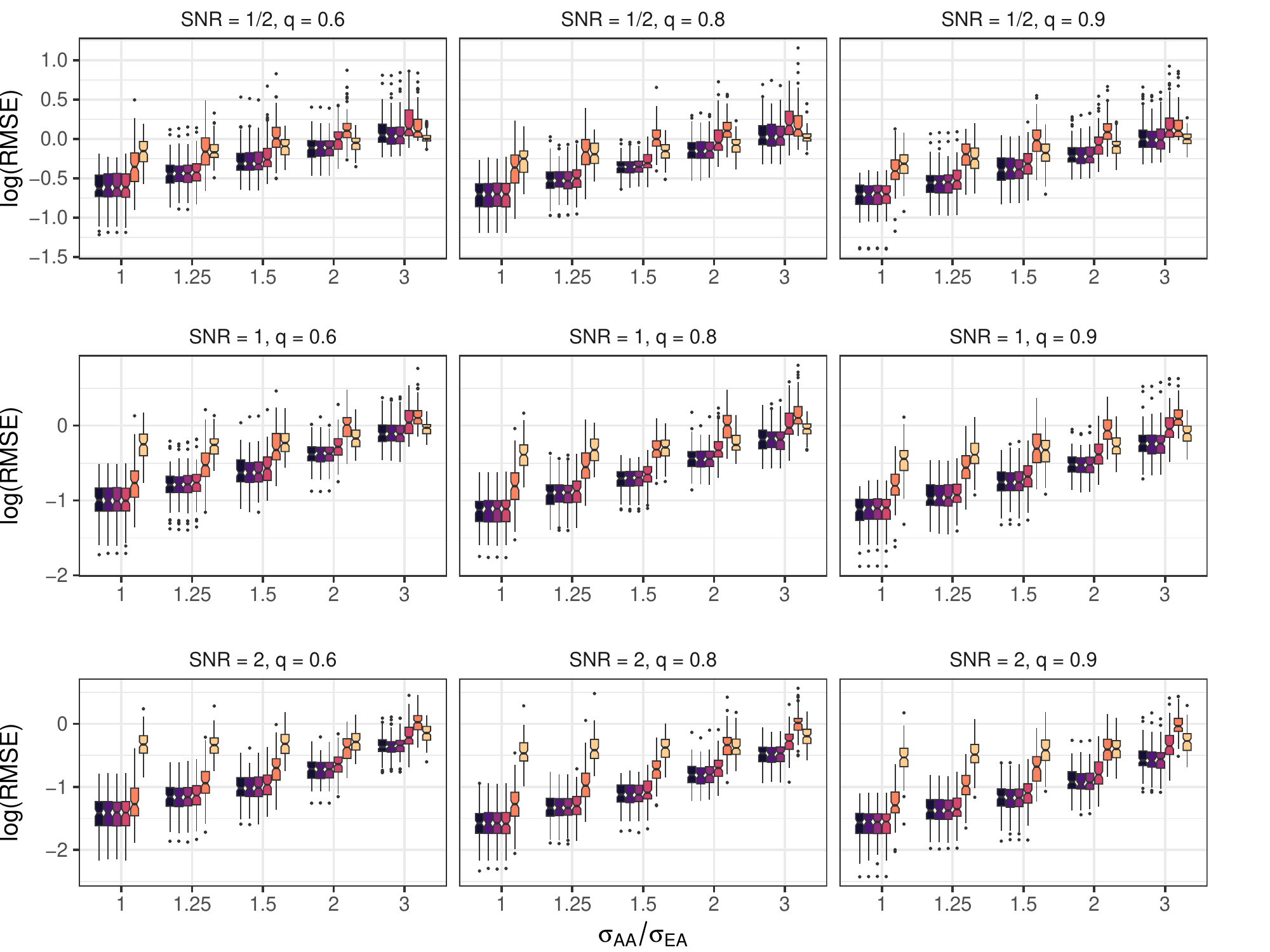}
\includegraphics[width=0.7\textwidth]{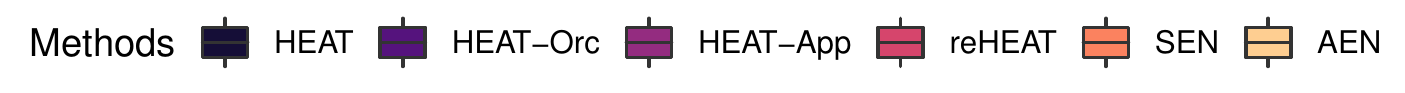}
\end{center}
\caption{Relative mean squared error results (on the log-scale) for 100 independent replications in each model setting when $n_{{\rm (AA)}} = 1027$. In each scenario, the SNR was used to fix the error variance in the EA population. }\label{fig:MSE}
\end{figure}

\begin{figure}[t]
\begin{center}
\includegraphics[width=0.9\textwidth]{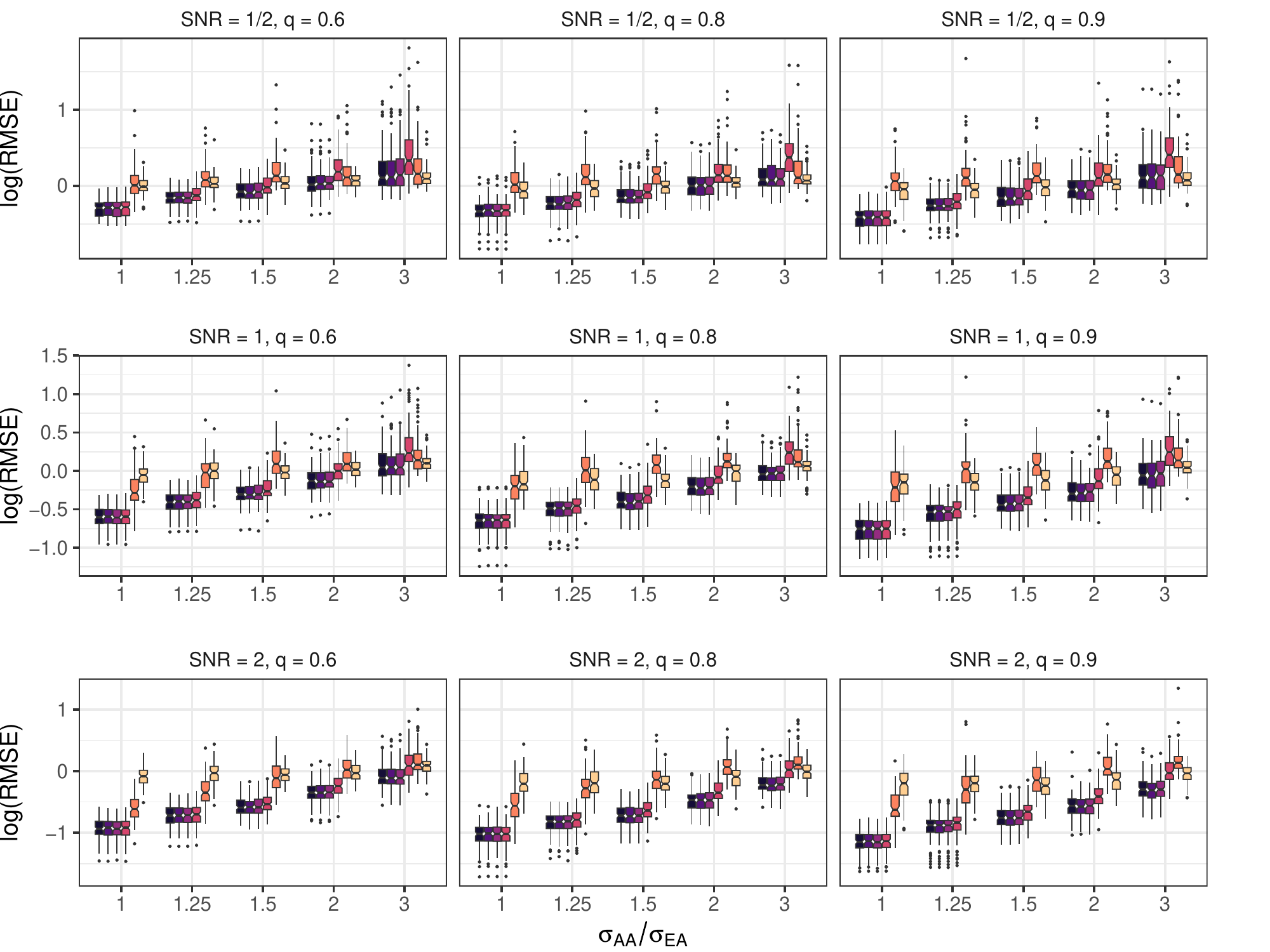}
\includegraphics[width=0.7\textwidth]{plots/Legend.pdf}
\end{center}
\caption{Relative mean squared error results (on the log-scale) for 100 independent replications in each model setting when $n_{{\rm (AA)}} = 433$. In each scenario, the SNR was used to fix the error variance in the EA population. }\label{fig:MSE_Halved}
\end{figure}

To assess the performance of an estimator $\widehat{\mbbeta}_{(\rm AA)}$, we measure relative mean squared error (RMSE), $\|\widehat{\mbbeta}_{\rm AA} - \mbbeta_{*(\rm AA)}\|_2^2/\|\mbbeta_{*(\rm AA)}\|_2^2$, and relative model error (RME), which we define as 
$\|\mbX_{\rm AA}(\widehat{\mbbeta}_{\rm AA} - \mbbeta_{*(\rm AA)})\|_2^2/\|\mbX_{\rm AA}\mbbeta_{*(\rm AA)}\|_2^2.$ 

\subsection{Results}
In Figure \ref{fig:MSE}, we display relative mean squared error results under varying SNR, $\sigma_{*(\rm AA)}/\sigma_{*(\rm EA)}$, and $q$.  In the top row of Figure \ref{fig:MSE}, where SNR $=1/2$, we see that \texttt{SEN} and \texttt{AEN} both perform worse than all \texttt{HEAT} variations when $\sigma_{*(\rm AA)}/\sigma_{*(\rm EA)}$ is less than or equal to 1.5. As $\sigma_{*(\rm AA)}/\sigma_{*(\rm EA)}$ increases, and the signal to noise ratio for the AA population decreases, we begin to see every method perform worse.  Once $\sigma_{*(\rm AA)}/\sigma_{*(\rm EA)} = 3$, \texttt{AEN} performs better than \texttt{SEN} and \texttt{reHEAT}. However, in all settings considered, the oracle, approximate, and exact versions of \texttt{HEAT}, perform as well or better than all the competitors.
Results do not change substantially across the considered values of $q$. The only method which is not affected by $q$ is \texttt{SEN}, which is unable to exploit shared pQTLs across populations. \textcolor{black}{In the Supplementary Material, we also include results under the same scenarios with $q = 1$: relative performances do not differ much from the case that $q = 0.9.$}


In the middle row of Figure \ref{fig:MSE}, we see a similar trend, though the difference between \texttt{SEN} and the \texttt{HEAT} variants is less substantial than in the top row. Notably, \texttt{AEN} never performs as well as the variants of \texttt{HEAT} other than \texttt{reHEAT}, which performs especially poorly when $\sigma_{*(\rm AA)}/\sigma_{*(\rm EA)} = 3$. In the bottom row of Figure \ref{fig:MSE}, the same trends are again observed, with even \texttt{reHEAT} outperforming \texttt{AEN} in most scenarios. 

Though the differences between \texttt{HEAT} and \texttt{reHEAT} are somewhat obscured by the variation across replications, \texttt{HEAT} significantly outperforms \texttt{reHEAT} even when $\sigma_{*(\rm AA)}/\sigma_{*(\rm EA)}$ is only slightly larger than one. To illustrate this point, we present the average ratio of mean squared errors of \texttt{HEAT} to \texttt{reHEAT} in Table 1 of the Supplementary Material. Here, we see that even when $\sigma_{*(\rm AA)}/\sigma_{*(\rm EA)} = 1.25$, \texttt{HEAT} provides a modest improvement over \texttt{reHEAT}. As $\sigma_{*(\rm AA)}/\sigma_{*(\rm EA)}$ increases, the improvement becomes more substantial. 



Turning our attention to the results in Figure \ref{fig:MSE_Halved}, wherein $n_{\rm (AA)} = 433,$ we see the same general trends as in Figure \ref{fig:MSE}. A notable difference, however, is that when $\sigma_{*(\rm AA)}/\sigma_{*(\rm EA)} = 3$ and SNR = $1/2$, \texttt{AEN} outperforms all competitors. This is due to the fact that with a smaller sample size, the large value of $\sigma_{*\rm (AA)}$ makes estimation of $\mbbeta_{*(\rm AA)}$ challenging. The method \texttt{AEN} is less affected by this since it does not distinguish between AA and EA. 

In the Supplementary Materials, we include relative mean squared error results for $\mbbeta_{*\rm (EA)}$, and relative model error results for $\mbbeta_{*(\rm AA)}$. To summarize briefly, in terms of EA mean squared error, the same general trends as in Figures \ref{fig:MSE} and \ref{fig:MSE_Halved} are observed. However, \texttt{reHEAT} tends to perform more closely to the other \texttt{HEAT} variants when  $\sigma_{*(\rm AA)}/\sigma_{*(\rm EA)} \leq 1.5$. In terms of model error, relative performances are similar to those in Figures \ref{fig:MSE} and \ref{fig:MSE_Halved}. \textcolor{black}{Also in the Supplementary Materials, we repeat a subset of simulation studies with SNPs from three different loci: the local/cis-SNP regions around the proteins ANGPTL3, SORT1, and GAS6 (see Table \ref{tab:ProteinsDiscovered}). The same relative performances were observed using these SNP sets. }
 
\section{Blood lipid proteome-wide association study}\label{sec:DataAnalysis}
\subsection{Data preparation}\label{subsec:dataPrep}

To begin, we obtained protein expression on 451 proteins from $n_{\rm AA} = 1033$ African American and $n_{\rm EA} = 866$ European American individuals, as measured by the OLINK targeted (866 European Americans and 351 African Americans) and explorer  (991 African Americans, 309 of which overlaps with Target) platforms (https://olink.com). \textcolor{black}{In this analysis, we defined an individual's group (AA or EA) based not on genetics, but on their self-identified race/ethnicity (SIRE). We discuss the reason for this in Section \ref{subsec:MetaXcan}.}

Next, following \cite{pietzner2021mapping} and \cite{zhang2022plasma}, we performed a two-step data processing. For each protein, we performed an inverse normal quantile transformation (separately in each of the two populations). We then regressed the transformed proteins against covariates and PEER factors to account for hidden factors that may confound the associations of interest  \citep{stegle2010bayesian}, and rank-inverse normalized the residuals, which we use as the response for building the prediction models in AA and EA. 

For each protein, we obtained cis/local-SNPs, defined as those SNPs within 1MB of the corresponding gene, with minor allele frequency $\geq$ 0.05 in either of the two populations. To avoid collinearity, we pruned highly correlated SNPs. Namely, starting with the SNP with highest variability, we removed all other SNPs with absolute correlation 0.95 or greater with this SNP. Then, moving on the SNP with the next highest variability (among those retained after the first pruning step), we repeated this process. This continued until no two SNPs had correlation greater than 0.95. \textcolor{black}{This pruning was performed separately in each population, and the union of the set of SNPs retained was used as the set of candidate SNPs (i.e., the same set of candidate SNPs were used for both EA and AA models). }

\textcolor{black}{For the vast majority of proteins, there are a substantial number of SNPs retained which do not vary at all across the EA population. In this scenario, we used the version of our method described in Section \ref{sec:AncestrySpecific}.}


 \begin{table}[t]
 \begin{center}

 \scalebox{0.9}{
\begin{tabular}{cccccc|cccc}
  \toprule
& & \multicolumn{4}{c|}{Ratio} & \multicolumn{4}{c}{Numerator}\\
&  & $Q_1$ & $Q_2$ & Avg. & $Q_3$ & $Q_1$ & $Q_2$ & Avg. & $Q_3$ \\ 
  \midrule
\multirow{2}{*}{$\bar{R}^2_{\rm HEAT}/\bar{R}^2_{\rm SEN}$}& AA   &0.977  &1.017&  1.203 & 1.114 & 0.031&  0.085 & 0.125 &  0.181\\
 & EA  & 0.986 & 1.016 & 1.168 & 1.122 & 0.029 &0.089& 0.147 & 0.230  \\ 
 \cline{2-10}
\multirow{2}{*}{$\bar{R}^2_{\rm HEAT-App}/\bar{R}^2_{\rm SEN}$} &AA  & 0.979 & 1.018 & 1.205 & 1.115  & 0.031  &0.084  &0.125 & 0.181 \\ 
&  EA &  0.998 & 1.030 & 1.242 & 1.185 & 0.031 & 0.090 & 0.149 & 0.232\\
 \midrule
\multirow{2}{*}{${q_{0.5}}(R^2_{\rm HEAT})/{q_{0.5}}(R^2_{\rm SEN})$} &AA   & 0.977 & 1.017 & 1.203 & 1.114  &  0.034 & 0.082  &0.126 & 0.181\\ 
  & EA & 0.979 & 1.018 & 1.205 & 1.115 & 0.032 & 0.089&  0.150 & 0.229 \\ 
  \cline{2-10}
\multirow{2}{*}{${q_{0.5}}(R^2_{\rm HEAT-App})/{q_{0.5}}(R^2_{\rm SEN})$} &  AA & 0.967 & 1.016 & 1.089 & 1.137  & 0.034 & 0.083  &0.126 & 0.182\\ 
&   EA & 0.979 & 1.034 & 1.144 & 1.179  & 0.035 & 0.089 & 0.152 & 0.232\\ 
   \bottomrule
\end{tabular}
}
\end{center}
\caption{Relative testing-set $R^2$ using our method (\texttt{HEAT}) versus separate elastic net estimators (\texttt{SEN}). $\bar{R}^2_{\rm HEAT}$ and ${q_{0.5}}(R^2_{\rm HEAT})$ denote mean and median of the testing-set $R^2$ over 25 independent training/testing splits within a protein. Summary statistics (quartiles [ Q1, Q2, and Q3] and average [Avg.]) are taken over proteins. The right columns provide the summary statistics for the numerator in each row. }\label{tab:results_WHI}
\end{table}

 \begin{figure}[t]
\begin{center}
\includegraphics[width=0.85\textwidth]{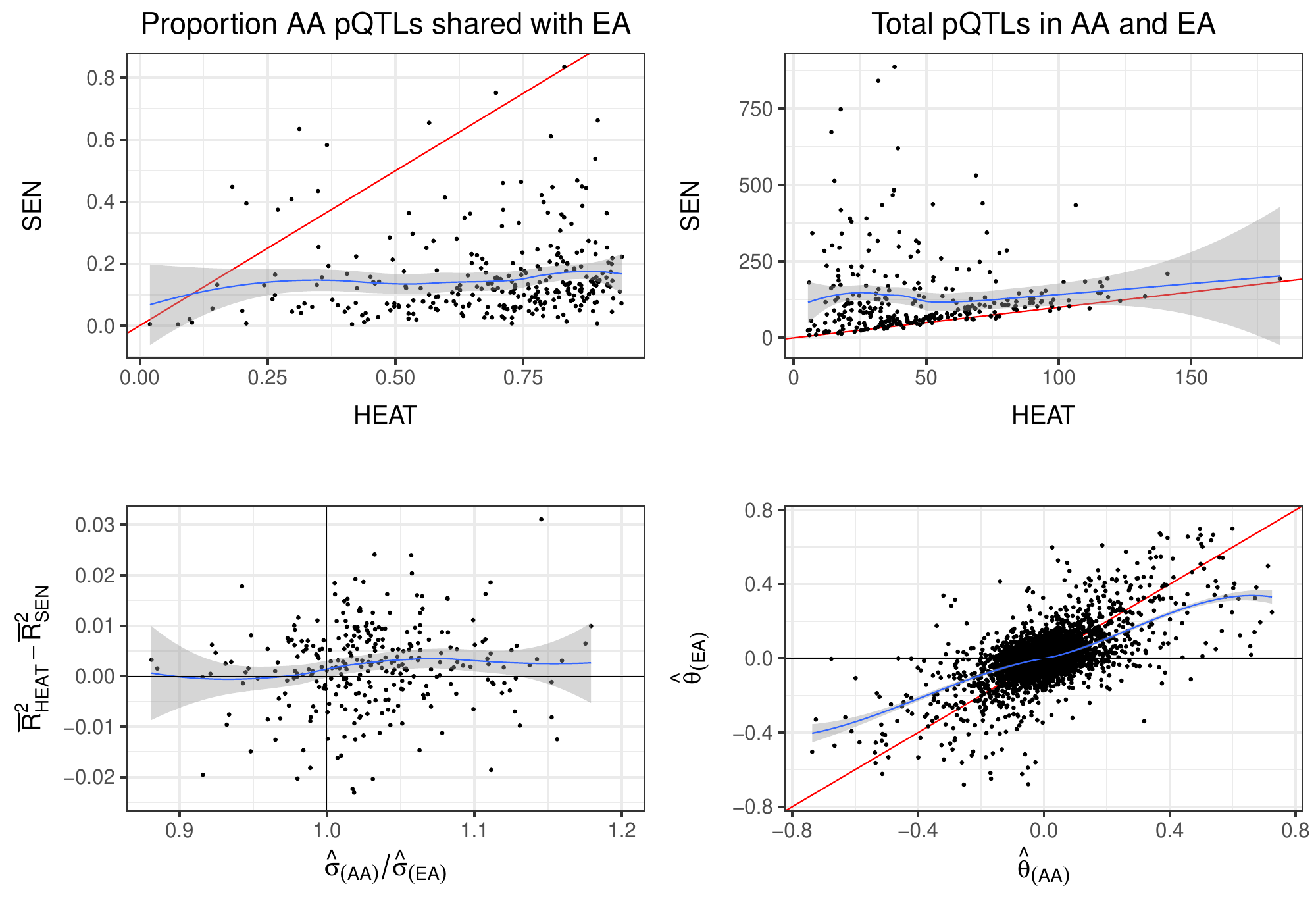}
\end{center}
\caption{Upper left: the proportions of AA pQTLs which are also estimated to be pQTLs in EA for \texttt{HEAT} and \texttt{SEN}. Upper right: the total number of estimated pQTLs in both AA and EA populations for \texttt{HEAT} and \texttt{SEN}. Lower left: the difference in average testing-set $R^2$ versus the estimated $\sigma_{*(\rm AA)}/\sigma_{*(\rm EA)}.$ We omit five points with horizontal axis value beyond 1.2 for improved display. Lower right: scatterplot of shared pQTL coefficients between EA and AA populations estimated by \texttt{HEAT}.}\label{fig:shared_pQTLs}
\end{figure}
\subsection{Protein expression prediction accuracy in WHI data}\label{subsec:AnalysisCompare}
We compared both the exact and approximate version of our method, \texttt{HEAT}, to fitting separate elastic net estimators, \texttt{SEN}, on data from AA and EA populations. We focused on \texttt{SEN} as our main competitor as this was the method used in \citet{zhang2022plasma}. We omitted comparisons to \texttt{AEN} because this method performed poorly in our simulation studies and would not allow us to discover pQTLs specific to the AA population. We also omitted comparisons to the other \texttt{HEAT} variant, \texttt{reHEAT}, because it never performed better than \texttt{HEAT} in our simulation studies. For 25 independent replications, we randomly split the data into training/testing sets of size 981/52 and 822/40 for AA and EA populations, respectively. For each protein, we calculated the means and medians of testing set $R^2$ for both AA and EA testing samples over 25 replicates. We then summarized the performance over all proteins by quartiles (Q1, Q2, Q3) and the average. 

For many of the proteins the average testing set $R^2$ is near zero, which would suggest that the SNPs do not explain any of the variability in the protein expression. For this reason, we present results only for proteins wherein at least one of the three methods under consideration had average testing set $R^2 \geq 0.01$, resulting in 289 proteins for AA and 297 proteins for EA. Results are displayed in Table \ref{tab:results_WHI}. Here, we can see that the exact \texttt{HEAT} and approximate \texttt{HEAT-App} performed similarly, and they were both better in terms of testing set prediction accuracy compared to separate fitting estimator \texttt{SEN}. 
For example, in the AA samples, the $R^2$ is on average around 20\% higher using our method \texttt{HEAT} than separate estimator \texttt{SEN}. 

Where \texttt{HEAT} clearly improved over separate elastic net estimators \texttt{SEN} was in terms of model complexity. Specifically, \texttt{HEAT} achieved improved prediction accuracy over \texttt{SEN} while using far smaller fitted models. In the majority of the proteins, less than 20\% of the AA pQTLs estimated by \texttt{SEN} were shared with the EA pQTLs. In contrast, in the majority of the proteins, more than 50\% of the AA pQTLs estimated by \texttt{HEAT} were shared with EA pQTLs (upper-left panel of Figure \ref{fig:shared_pQTLs}).  This degree of pQTLs sharing across these two populations was expected: for example, in the analysis of \citet{zhang2022plasma}, they found that of 1618 pQTLs identified in AA, 1447 (89.4\%) were also pQTLs in EA (though they were defining pQTLs marginally, rather than jointly).

In the upper-right panel of Figure  \ref{fig:shared_pQTLs}, we display the total number of pQTLs estimated for both EA and AA. Because the majority of pQTLs estimated using \texttt{HEAT} are shared across populations, it is not surprising that \texttt{HEAT} had fewer SNPs to be pQTLs than \texttt{SEN}. The fact that \texttt{SEN} selected such a large number of additional SNPs as pQTLs to achieve a worse prediction accuracy raises concerns about the reliability of downstream association testing of genetically predicted proteomic expressions based on the \texttt{SEN} fitted model.  In the bottom left panel of Figure \ref{fig:shared_pQTLs}, we see that, roughly, as $\hat\sigma_{\rm AA}/\hat\sigma_{\rm EA}$ increases, it appears that our method tends to slightly outperform \texttt{SEN} in terms of prediction accuracy. In the bottom right panel, we see that in general, the coefficients for AA estimated by our method had slightly larger magnitude than the coefficients for EA. This may be influenced by the fact that our dataset has more AA than EA subjects, so the penalty has a stronger effect on EA coefficients. 
\begin{table}[t]
\begin{center}
\scalebox{0.8}{\begin{tabular}{r|ccccc}
\toprule
Proteins & HDL & LDL & logTG & nonHDL & TC\\
\midrule
LILRB5 & \checkmark & & & & \\
LPL& \checkmark &  & X & & \\
PLTP& \checkmark & & \checkmark& & \\
LILRB1 & \checkmark & & & \\
ANGPTL3& \checkmark & \checkmark &X & X &X \\
TNFSF14& & \checkmark & && \checkmark\\
IGFBP6 & &\checkmark & & & \checkmark\\
GAS6 & & \checkmark&\checkmark & & \checkmark\\
PLAUR & & \checkmark & & & \checkmark\\
PCSK9 & & \checkmark& & & X\\
SORT1 & & \checkmark& &\checkmark &X \\
CCL24 & & & & & \checkmark \\
HSPB1& &  & & & \checkmark\\
HLA-E& &  & &  & \checkmark\\
IL1RN& &  & &  & \checkmark\\
\bottomrule
\end{tabular}}
\caption{Proteins surpassing the Benjamini-Hochberg corrected p-value threshold 0.05. Associations denoted with an X are those that had a pQTL surpassing the genome-wide significance threshold $5\times10^{-8}$ for association with the trait.  }\label{tab:ProteinsDiscovered}
\end{center}
\end{table}

\subsection{Association analysis of genetically predicted protein expression with blood lipid traits}\label{subsec:MetaXcan}
We performed the group-specific association analysis of genetically predicted protein expression based on our fitted models with blood lipid traits with GWAS summary statistics using MetaXcan \citep{barbeira2018exploring}. 
We use coefficients estimated by \texttt{HEAT}, which were obtained as described in Section \ref{subsec:AnalysisCompare}, except that we used the entire dataset for model fitting. 
Focusing only on proteins whose average testing set $R^2 \geq .01$ using our method, we tested 286 proteins in total. To perform the MetaXcan, we obtained LD matrices using the AA individual level genotyping data from the WHI. Then, we tested the relationship between genetically predicted protein expression and five blood lipid traits: LDL, HDL, TG, TC, and nonHDL \citep{graham2021power}. The summary statistics were downloaded from the Global Lipids Genetics Consortium (GLCC, http://csg.sph.umich.edu/willer/public/glgc-lipids2021/). These summary statistics are from 99,432 individuals of African ancestry. \textcolor{black}{Consistent with our ancestral group definition in Section \ref{subsec:dataPrep}, GLCC also defines ancestry based on SIRE.}

In Table \ref{tab:ProteinsDiscovered}, we present the set of proteins identified as significant at the 0.05 level after Benjamini-Hochberg correction in each of the five traits. Out of the 15 proteins, four proteins included a pQTL surpassing the conventional genome-wide significance threshold $5 \times 10^{-8}$ for the association with the trait. It is reassuring that a number of proteins including LPL, ANGPTL3, PCSK9, PLTP, and SORT1 are well-known for their genetic determinant being associated with lipid traits across racial and ethnic groups \citep{paththinige2017genetic, musunuru2012multi, graham2021power, selvaraj2022whole}. Other proteins may offer novel insights. In particular, PLAUR, which encodes the receptor for urokinase plasminogen activator, plays an important role in blood coagulation and inflammation as well as cardiovascular disease risk \citep{mahmood2018multifaceted}. PLAUR has been reported to have several strongly ancestry-differentiated coding and regulatory cis-pQTL variants that are relatively common in Blacks \citep{olson2021soluble}.    

We performed the same association test using the weights estimated by \texttt{SEN} for the AA population. A similar total number of significant associations were discovered, though the two are not comparable since different sets of proteins were tested. The significant findings from \texttt{SEN} weights were slightly more often attributed to a single SNP genotype which reached genome-wide significance included in the protein prediction model than our method \texttt{HEAT}.  To demonstrate the importance of ancestry-specific weights, we also performed the association testing using the set of weights estimated by \texttt{SEN} on the EA data. Using these weights, we discovered only 19 associations---far fewer than the 30 displayed in Table \ref{tab:ProteinsDiscovered}.  This further reinforced the observation that genetic prediction models were often not portable across ancestral groups \citep{bhattacharya2022best}. 

\textcolor{black}{Before concluding this section, we note that our results may be somewhat biased by the fact that our fitted models for protein expression in AA were trained on data from women only, whereas the GLCC summary statistics were obtained from a cohort including both men and women. }

\section{Discussion}
\textcolor{black}{A referee raised an important point about African ancestry-specific modeling: often, the data consists of admixed African ancestry individuals with varying proportions of their genomes inherited from European ancestors. Self-identified race/ethnicity, a social construct, has been commonly used to group people, including admixed individuals, in genome-wide association studies. However, for genetic studies, it would be natural to instead define groups according to genetically inferred ancestry \citep[e.g., as in ][]{oak2020ancestry}. As suggested by a referee, one could choose a threshold according to the RFmix proportion \citep{maples2013rfmix} or according to ADMIXTURE proportions \citep{alexander2009fast}. For example, \citet{oak2020ancestry} assign individuals to an ``admixed'' group if they have two ancestral proportions greater than 0.20 estimated by ADMIXTURE. Many other approaches for ancestry inference exist \citep[Section 3,][]{tan2023strategies} and could be used. Notably, SIRE and genetically inferred ancestry often coincide \citep[e.g., see extended data Figure 10 of ][]{zhang2022plasma}, or they may be harmonized in order to define discrete ancestral groups \citep{fang2019harmonizing}.  }

\textcolor{black}{To address admixture directly, it may be useful to devise a method that considers ancestry as continuous rather than discrete.  For example, we could postulate that the random protein expression for the $i$th individual, $Y_i$, with a local African ancestry proportion $\pi_i \in [0,1]$ and SNP genotypes $x_i \in \mathbb{R}^p$ is given by 
$ Y_i = x_i^\top b_*(\pi_i) + \epsilon_i(\pi_i)$
where $b_*:[0,1] \to \mathbb{R}^p$ is an unknown, component-wise smooth function \citep{hastie1993varying} and $\epsilon_i(\pi_i)$ is a random error whose distribution depends on $\pi_i$ (e.g., $\epsilon_i(\pi_i) \sim {\rm N}\{0, \sigma_1^2 \pi_i + \sigma_2^2 (1- \pi_i)\}$). The vector $b_*(0) \in \mathbb{R}^p$, for example, is the coefficient vector for an individual with no local African ancestry. Alternatively, one could directly use the inferred local ancestry using methods such as RFmix \citep{maples2013rfmix}, HAPMIX \citep{salter2019fine}, and FLARE \citep{browning2023fast}. In this scenario, the group would be defined locally at each SNP for each individual, which could complicate regression modeling of protein expression based on cis/local SNP genotypes. Future research on the best approach to handling admixed individuals in this context is warranted.  }

\textcolor{black}{
Finally, while our method borrows information across populations, there is an opportunity for further efficiency gains through ``seemingly unrelated regressions"-type joint modeling \citep{zellner1962efficient} of all proteins simultaneously. In brief, this approach assumes that the errors for each protein's regression are correlated within a population. Modeling the covariance between these errors can improve efficiency. We discuss this in detail in the Supplementary Material. }


\bibliography{Heterogeneous_pQTL}

\end{document}